# Directional emission of a deterministically fabricated quantum dot – Bragg reflection multi-mode waveguide system

*Paweł Mrowiński,[1] Peter Schnauber,[1] Philipp Gutsche,[2] Arsenty Kaganskiy,[1] Johannes Schall,[1] Sven Burger,[2] Sven Rodt[1] and Stephan Reitzenstein[1,\*]*

[1] Institut für Festkörperphysik, Technische Universität Berlin, Hardenbergstraße 36, 10623 Berlin, Germany

[2] Zuse Institute Berlin, Takustraße 7, 14195 Berlin, Germany

ABSTRACT

We report on the experimental study and numerical analysis of chiral light-matter coupling in deterministically fabricated quantum dot (QD) waveguide structures. We apply in-situ electron beam lithography to deterministically integrate single InGaAs/GaAs QDs into GaAs-DBR waveguides to systematically explore the dependence of chiral coupling on the position of the QD inside the waveguide. By a series of micro-photoluminescence measurements, we determine the directionality contrast of emission into left and right traveling waveguide modes revealing a maximum of 0.93 for highly off-center QDs and an oscillatory dependence of this contrast on the QD position. In numerical simulations we obtain insight into chiral light–matter coupling by computing the light field emitted by a circularly polarized source and its overlap with multiple guided modes of the structure, which enables us to calculate directional β-factors for the quantum emitters. The calculated dependence of the directionality on the off-center QD position is in good agreement with the experimental data. It confirms the control of chiral effects in deterministically fabricated QD-waveguide systems with high potential for future non-reciprocal on-chip systems required for quantum information processing.

**INRODUCTION**

On-chip quantum optics and photonics is emerging as a highly attractive platform to boost quantum information processing in the near future and to promote the realization of quantum nodes in large-scale quantum networks.[1] One of the important functionalities of on-chip nanophotonic systems is the directional emission and non-reciprocal scattering of single photons, which is required for the implementation of quantum optical diodes, gates or photonic transistors.[2–4] Non-reciprocity and single-photon routing can be achieved by applying chiral light-matter interaction between a two-level system such as a nitrogen vacancy center in diamond or a quantum dot (QD) and a propagating mode in a waveguide structure. Here, chiral coupling occurs due to the spin-momentum locking effect which is related to the time-reversal symmetry of the Maxwell's equations. This locking effect is the inherent link between the propagation direction and the transverse spin angular momentum (SAM) of the propagating mode.[5,6] In this context, the transverse SAM is related to the non-zero longitudinal electric field component that can appear in focused Gaussian beams, evanescent waves on dielectric interfaces, or in transversally confined modes as for guided modes in waveguides.[5]

Self-assembled semiconductor QDs are excellent two-level emitters in the solid state. They have been very successfully implemented as single-photon emitters both for free-space[7–12] and on-chip propagation.[2,3,13–17] Emission of QDs is based on optical transitions of $\pm 1$ spin states, which are distinguishable in circular polarization or spectrally due to the Zeeman effect if an external magnetic field is applied.[18] In order to obtain directional spontaneous emission of the QD into propagating modes of a waveguide, both the spatial and polarization matching condition must be fulfilled. This requires that the QD must be located in the chiral position of the respective mode, i.e. in the position of a circular polarization singularity defined by a Stokes vector component V =

$\mathrm{Im}(\mathbf{E}(\mathbf{r}) \times \mathbf{E}(\mathbf{r})^*) = \pm 1$, where $\mathbf{E}(\mathbf{r})$ is the electric field vector at position $\mathbf{r}$. Recently it has been shown that chiral positions can be found in propagating TE modes of nanobeam waveguides,[3] or in 2D photonic crystal waveguides with glide geometry[2] for off-center in-plane positions, while the linear polarization state dominates in the center of the waveguide. Chiral points in semiconductor ridge-type WGs that are quite robust and whose fabrication can be scaled-up easily, have not been shown until now.

In waveguide structures that support transversally confined propagating modes, spontaneous emission of the emitter's $\pm 1$ exciton/trion state can exhibit high directionality described by the associated $\beta_\pm$-factor. The directional $\beta_\pm$-factor quantifies the probability to emit a photon into the propagating mode in a certain direction relative to the total decay rate, and it is given by

$$\beta_\pm = \frac{\gamma_\pm}{\gamma_+ + \gamma_- + \Gamma_{out}}, \qquad (1)$$

where $\gamma_\pm$ is the emission rate into the guided mode propagating along the left (-) and right (+) waveguide axis. The coefficient $\gamma_\pm$ is proportional to the product of the transition dipole moment $d_{QE}$ and the electric field intensity of the mode at the dipole position $\mathbf{E}(r_{QE})$, $[|\mathrm{d}_{QE} \cdot \mathrm{E}(\mathrm{r}_{QE})|^2]$, while $\Gamma_{out}$ denotes the residual recombination rate by all other channels including losses due to non-guided modes and non-radiative intrinsic recombination.[6] One refers to directional coupling when non-symmetric emission occurs in a waveguide with $\beta_- \neq \beta_+$. In contrast, chiral coupling is defined as a fully directional propagation ($\beta_- = 0$ and $0 < \beta_+ \leq 1$, or vice versa), which in the ideal case of a fully deterministic light-matter interface also requires $\beta_+ \cong 1$.[19,20] In order to characterize the directional coupling one can use the relative directionality contrast[3,21] given by

$$C = \frac{\beta_- - \beta_+}{\beta_- + \beta_+} \qquad (2)$$

so that σ± chiral points are characterized by C = ±1. Alternatively, directional coupling can be described by the directionality factor[22] $\beta_-/(\beta_- + \beta_+)$ for σ+ (σ−) reflected by a probability of 0 (1). Besides the directional emission effect, the coupling regime also affects the transmission, reflection and absorption coefficients for the photons propagating along the waveguide[19] and, based on that, non-reciprocal devices such as optical circulators for single-photons can be realized.[19]

For a controlled study of chiral light-matter coupling and for the later use of directionality it is crucial to precisely control the position of the quantum emitter in the WG with an accuracy on the order of a few tens of nm. This requirement is highly challenging with respect to the fabrication technology and can hardly be fulfilled using standard non-deterministic nanotechnology methods. In-situ deterministic fabrication techniques, on the other hand side, are very attractive as they allow for the selection of QDs with certain optical properties and their spatially controlled integration in nanophotonic structures and devices. Several concepts for deterministic QD device integration have been developed in recent years. This includes the site-positioned growth of QDs,[23,24] low-temperature optical in-situ lithography[25,26] and marker-based pre-characterization and lithography,[3,27,28] which has already been applied to QD chiral WG devices.[2,3]

Here we report on the directional coupling of QD emission in deterministically fabricated ridge-type distributed-Bragg-reflection waveguide (DBR WG) devices. Applying deterministic in-situ electron-beam lithography (EBL) we are able to pre-select single QDs and integrate them into WGs at a chosen position[13] with an accuracy of 30-40 nm.[29] We employ the DBR ridge WG geometry serving a threefold use. Firstly, it allows for a straightforward almost 100% yield fabrication when using in-situ EBL. Secondly, the bottom DBR increases light emission normal to the surface enhancing the QD positioning accuracy in the deterministic fabrication process.

Thirdly, the DBR enables grating-like vertical outcouplers in a GaAs-AlGaAs non-underetched sample design. Our deterministic fabrication method provides the unique opportunity to systematically explore chiral light-matter coupling as a function of the displacement (Δx) of our two-level emitter from the center of the WG. The deterministic QD integration also helps to better understand the underlying physics of chiral light-matter coupling by comparing the experimental results to predictions obtained by numerical modelling based on the finite-element method. This technology platform and the detailed understanding of the optical properties will be key components for the realization of more complex quantum photonics circuits based on chiral light-matter coupling.

## DETERMINISTIC FABRICATION OF QUANTUM DOT WAVEGUIDES USING IN-SITU ELECTRON-BEAM LITHOGRAPHY

Figure 1a) illustrates schematically the QD-WG system realized and studied in our work. The structure design leads to directional emission into propagating WG modes depending on the spin state of the photon emitted by a deterministically integrated QD. Grating outcouplers at both ends of the WG facilitate optical access in normal direction to the sample surface. By applying external magnetic fields in Faraday configuration, Zeeman splitting occurs which allows one to spectrally distinguish between $S = \pm 1$ states of charged excitonic complexes of the QD. In Fig. 1b) the layer structure of the GaAs-DBR WG cross-section is presented. The sample was grown by metal-organic chemical vapor deposition (MOCVD) on (100) GaAs substrate. A 300 nm thick GaAs buffer is followed by 23 mirror pairs of AlGaAs/GaAs forming a DBR whose stopband is centered at 930 nm. The DBR constitutes the WG cladding layer. The DBR is followed by GaAs layer with a thickness of 230 nm which includes a single layer of InGaAs QDs with a density of $10^8$ cm$^{-2}$. This sample design supports guided TE-like modes inside the GaAs λ-layer in 800 nm deep etched ridge waveguides, for which a guided-mode confinement is expected for WG widths of more than 160 nm, and single-mode behavior is valid up to 330 nm.

We use in-situ EBL for WG fabrication[13] to realize a family of ridge QD-WG devices with D-shaped vertical outcouplers with an overall length of 50 µm and a width of 800 – 850 nm for which the lateral position of the QD (i.e. in perpendicular direction to the WG, 'Δx' in Fig. 1b)) was varied from -350 to 350 nm. The in-situ EBL localization and patterning step is schematically visualized in Fig. 2a). In the first step, we locate and pre-characterize single QDs emitting around 930 nm spectral range inside the sample, which has been spin-coated with AR-P 6200.04 dual-tone EBL resist and cooled down to 10 K.[30] Scanning the sample at an area electron dose of 10 mC/cm² and taking cathodoluminescence (CL) spectra at the same time (500 nm grid size, 50 ms exposure time, 0.5 nA beam-current), we acquire CL maps in which we can locate multiple spectrally and spatially isolated QDs with an accuracy of ~20 nm with each scan, without crosslinking the resist.[31] The λ-cavity DBR reflector design increases the amount of light captured by the 0.8 NA elliptical mirror inside the sample chamber as compared to non-DBR sample structures, which guarantees maximum positioning accuracy for every structure. Immediately after the localization step, we use grey-scale lithography to pattern a proximity-corrected WG etch mask with respect to the QD location, resulting in an overall integration accuracy of ~34 nm.[32] The EBL patterns are transferred into the sample through anisotropic dry etching.

The etch mask is a proximity-effect corrected grey scale pattern, which has been prepared before cool-down. Through proximity effect measurements the radial dose distribution inside the 10 K temperature resist under electron beam irradiation film is determined.[33] The proximity-correction is based on a home-made induced-electron-dose simulator that uses this proximity function and the resist contrast curve to calculate the expected etched structure from a given EBL pattern taking into account the QD localization exposure step. Subsequently, the simulator iteratively corrects and re-simulates the pattern until it converges to a grey-scale mask that will yield in the target structure. Our home-made EBL pattern processor has been extended such that the fracturing and patterning of patterns with a grey-scale resolution of 6 bit and pattern resolution of 5 nm is handled within a few seconds. At the operating voltage of 20 kV and at a temperature of 10 K, the employed EBL resist shows a pronounced long-range proximity contribution, which significantly reduces the dose-tolerance window for EBL patterns that contain voids or grating-like features as in the vertical outcoupler. By applying the proximity correction, a single calibration step allows for an 'as-designed' structure manufacturing process with 100% yield. Thereby, this is the first report on the successful application of proximity correction in low-temperature in-situ EBL.

Fig. 2b) shows a scanning electron microscopy (SEM) image of a deterministically fabricated QD-WG device. At both ends of the WG we patterned D-shaped vertical outcouplers as shown in fig. 2c) to enhance the collection efficiency by a microscope objective (NA = 0.4, magnification x20) in normal direction. The D-shaped outcoupler scatters the light guided in the TE modes towards the bottom side of the structure due to the low refractive index contrast with the cladding bottom part (no symmetrical scattering towards top and bottom as in nanobeam WG grating couplers). Subsequently, the light is reflected by the DBR mirror and directed towards the top. This DBR-sample D-shaped coupler geometry enables towards-the-top light coupling in high index cladding ridge WGs and is to our knowledge the first demonstration of such a structure.

**SPECTROSCOPIC STUDY OF CHIRAL LIGHT-MATTER COUPLING IN DETERMINISTIC QUANTUM DOT WAVEGUIDES**

The fabricated samples were characterized in a high-resolution two-beam micro-photoluminescence (µPL) setup. It includes a confocal arrangement for spatial selection of emission from the WG-outcoupler under simultaneous laser excitation at the QD position in the center of the WG. The sample was cooled down to 6-8K in either in standard Helium flow cryostat, or in a magneto-optical Helium flow cryostat providing magnetic fields of up to 5 T. The QDs were excited by a diode laser at 787 nm. For detection we used 750 mm focal length spectrometers with liquid nitrogen-cooled or Peltier-cooled Si-CCD cameras with a spectral resolution of ~25 µeV. The polarization of emission was analyzed by achromatic half- and quarter-wave plates in front of a linear polarizing filter in the detection path. The confocal setup is realized by an aperture of 200 µm between two achromatic lenses with focal lengths of 10 cm. Suitable alignment of the aperture allows us to distinguish between QD emission at the center of the waveguide and light guided through the WG and finally scattered out of one of the outcouplers within the effective field of view (~50 µm) of the single microscope objective used for excitation and detection. The outcoupled QD emission is further directed to the monochromator with a scanning mirror used for raster scanning on the sample surface.

In our spectroscopic studies we first performed a basic characterization of QD excitonic states directly from the WG center. Figure 3a) depicts low and high excitation power spectra showing characteristic excitonic states such as the neutral exciton (X), charged excitons (X+, X-) and higher order excitonic complexes on the low energy side labeled as XX- and X-* possibly involving the p-shell states which is typical for the high excitation power regime.[34] Moreover, the linearly polarized emission scan presented in Fig. 3b) allows us to identify a fine structure splitting of 70 µeV for the X state and confirms the charged exciton character of adjacent lines due to lack of energetic splitting in the linear polarization basis.

Figure 3c) presents magneto-optical X- emission spectra of a deterministic QD-waveguide system with an off-center QD position of Δx = 300 nm and a WG width of (850 ± 20) nm. The upper panel shows the magnetic field dependent Zeeman splitting of circularly polarized right ($\sigma^+$) and left ($\sigma^-$) polarized states recorded directly from the QD position in the axial center of the WG. The PL intensity is similar for both states. Importantly, in case of the propagated emission detected at the left and right outcoupling elements (middle and lower panel in Fig. 3c) we observe almost perfect inhibition of one emission state revealing a directionality contrast of about 0.93 for B > 0.5 T, which indicates almost fully chiral coupling in this deterministic QD-WG system.

**NUMERICAL STUDIES OF DIRECTIONAL EMISSION**

To better understand the underlying physics and to support the experimental findings we studied chiral coupling and directional emission in QD-WGs by using numerical simulations based on Maxwell's equations. For solving the two- and three-dimensional models related to the propagating eigenmodes (2D) and to the dipole emission (3D) we use a higher-order finite-element method (FEM), implemented in the solver JCMsuite.[35] First, to understand the properties of the waveguide, we compute the propagating eigenmodes for a waveguide width of 850 nm and a height of 800 nm at 930 nm wavelength. Next, we introduce the QD emission to the numerical model by placing an in-plane circularly polarized dipole source in the WG. The directional $\beta_\pm$-factors and the contrast are then deduced from the near field by overlap integrals between the emitted field distributions and the propagating waveguide modes for both sides.

The results yield multiple propagating TE and TM modes localized in the GaAs top layer containing the QDs. [more details can be found in the Supplementary Information] The electric field intensity profiles of the fundamental TE mode and a 3[rd] higher order TE mode are presented in Fig. 4a), and in panel b) the in-plane degree of circular polarization given by a Stokes vector component V is shown, respectively. While the V distribution is valuable to identify the positions of high directionality for the in-plane circularly polarized dipole, we observe that the fundamental TE mode is localized in the center in contrast to the high V localized one close to the waveguide edge. This results in a substantially low directional $\beta_{\pm}$-factor of less than 1% for dipoles located at high V positions. However, in case of higher order propagating TE modes confined in the waveguide, as for example of the 3$^{rd}$ TE mode, both high intensity and high V can be expected in this region in combination with directional $\beta_{\pm}$-factors on the order of 5 % and V as high as 0.95 (see Supplementary Information, Fig. 2b). Due to the multiple guided modes in such DBR waveguide structures, an oscillatory behavior of the directional $\beta_{\pm}$-factors for a dipole emitter as a function of the spatial alignment Δx is expected, as presented in Fig. 4c). Our calculations predict that the total directional $\beta_-$ factor for a dipole position of Δx = 375 nm is about 20%, while the value of $\beta_+$ is around 1%, which gives a contrast as high as C = 0.89 or alternatively, a directionality factor $\beta_-/(\beta_- + \beta_+)$ = 0.93 in this case. A visualization of a typical resulting 3D field distribution for highly off-center dipole position is also shown in the inset of Fig 4c).

**DISCUSSION AND CONCLUSIONS**

Following both the experimental results and theoretical predictions presented so far, we measured the statistics of directionality contrast in this system for approximately 60 deterministically integrated QD-WG structures, focusing on charged emission of excitonic complexes (X+/X-), in dependence of the lateral QD position Δx in a waveguide ranging from -350 to 350 nm. The statistical experimental results are plotted in Fig. 5, where we identify signatures of oscillations in the directionality contrast which we interpreted in terms of an x-position dependent coupling to higher-order guided modes confined in the waveguide. The corresponding superimposed calculated dependence for a WG width of 850 nm is in good agreement with the data, regardless to the rather large standard deviation which is attributed to the accuracy of the WG width ($\pm 20$ nm), QD position ($\pm 34$ nm) and directionality measurement ($\pm 0.05$). In case of the utmost QD positions, the average directionality contrast is on the level of 0.33 which fits to the calculated dependence, nevertheless we should also emphasize that we measured sporadic values of maximally 0.93 which is close to the calculated limit of ~0.89.

In conclusion, our experimental and theoretical study of directional emission in deterministically fabricated QD waveguide devices demonstrates an attractive concept to obtain well-controlled non-reciprocal systems for on-chip quantum nanophotonics. Using proximity-corrected grey scale in-situ e-beam lithography for QD selection and their precise integration into WG structures provides high accuracy of the QD alignment to maximize chiral light-matter coupling by choosing the most suitable off-center QD positions. A D-shaped vertical outcoupler in combination with the DBR cladding layer enables the collection of QD emission normal to the sample surface. By placing single QDs 350 nm away from the center of an 850 nm wide waveguide, we observed highly directional propagation of in-plane circularly polarized emission from a charged excitonic state with a contrast of up to 0.93. Our deterministic device processing enabled systematic studies on the position dependence of directionality. Experimental results indicate an oscillatory variation of the directionality as a function of the QD displacement which is quantitatively confirmed by numeric results on the directional $\beta_{\pm}$-factors obtained by FEM modelling of the coupled QD-WG system in terms of the overlap of the multiple propagating eigenmodes and the circularly polarized dipole source. Our results provide important insight into the position dependent chiral light-matter coupling behavior of QD-waveguides which we studied for the first time in a deterministic and systematic way. Being able to precisely control the position of our quantum emitters in the WG our fabrication concept based on in-situ EBL can pave the way for more advanced quantum photonic circuits taking full advantage of the great opportunities arising from chiral-coupling enabled on-chip single-photon routing.

**FIGURES**

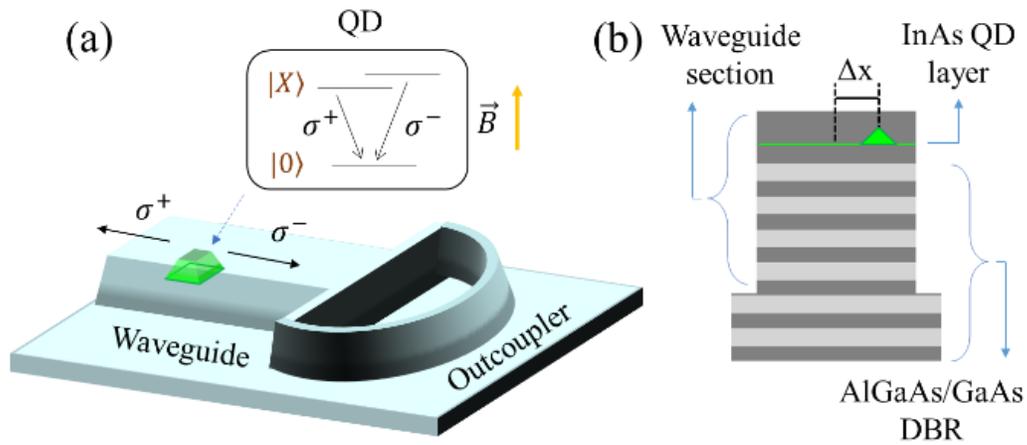

**Figure 1.** a) Schematic view of the QD-ridge-waveguide structure for the investigation of chiral light-matter coupling and the directional emission of single photons. b) Cross-section of the waveguide structure with a low-density InGaAs/GaAs QD layer above an AlGaAs/GaAs distributed Bragg reflector. The "Δx" represents the QD displacement with respect to the WG center.

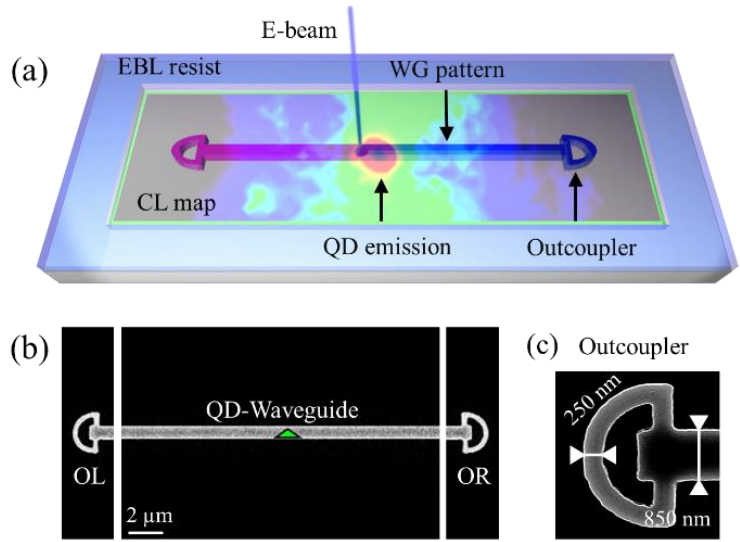

**Figure 2.** a) Illustration of the in-situ EBL lithography used for the fabrication of the deterministic QD-waveguide structures presenting an overlaid low-temperature cathodoluminescence (CL) map showing a 2D CL intensity profile (linear scaling, $\lambda$ = 930 nm) of emission of a pre-selected and integrated QD – see text. b) Scanning electron microscope (SEM) image of a processed QD-waveguide structure with outcouplers (OL and OR) c) SEM image of an outcoupler for enhanced scattering of emission from the end facette of the waveguide to the collection optics normal to the sample surface.

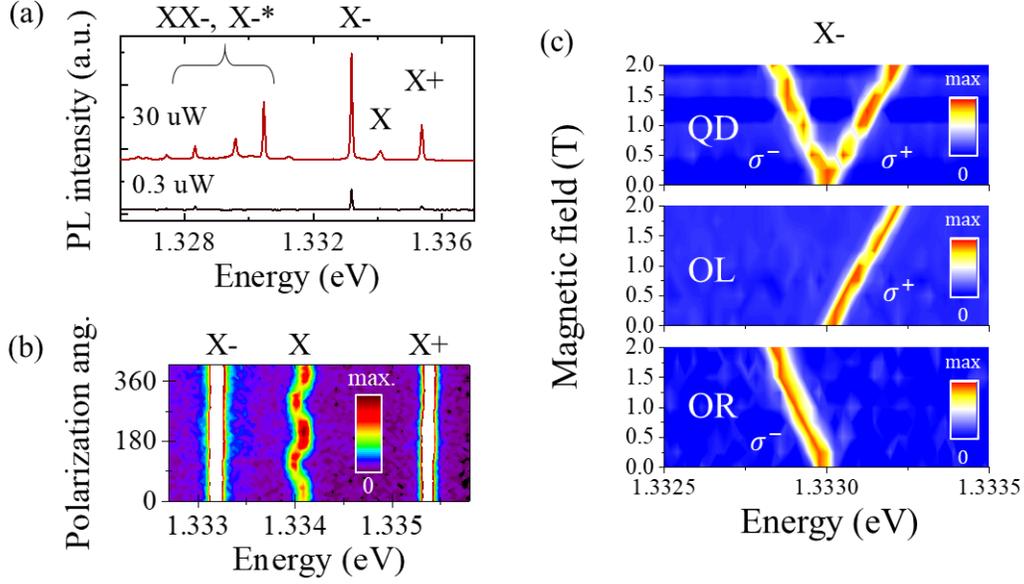

**Figure 3.** a) Low/high excitation power micro-photoluminescence spectra of a QD inside a deterministically fabricated QD-WG device. b) Linear polarization scan showing bright charged excitons (X+, X-) and neutral exciton (X) emission characterized by a fine structure splitting of about 70 µeV. c) QD (X-) emission split by an external magnetic field for detection of $\sigma^{\pm}$ polarized states observed from the waveguide center (QD position) and from both right and left outcouplers (OR and OL) showing a directionality contrast of up to 0.93 (see eq. 2). Here the lateral QD position is $\Delta x = 350$ nm and the WG width is 850 nm.

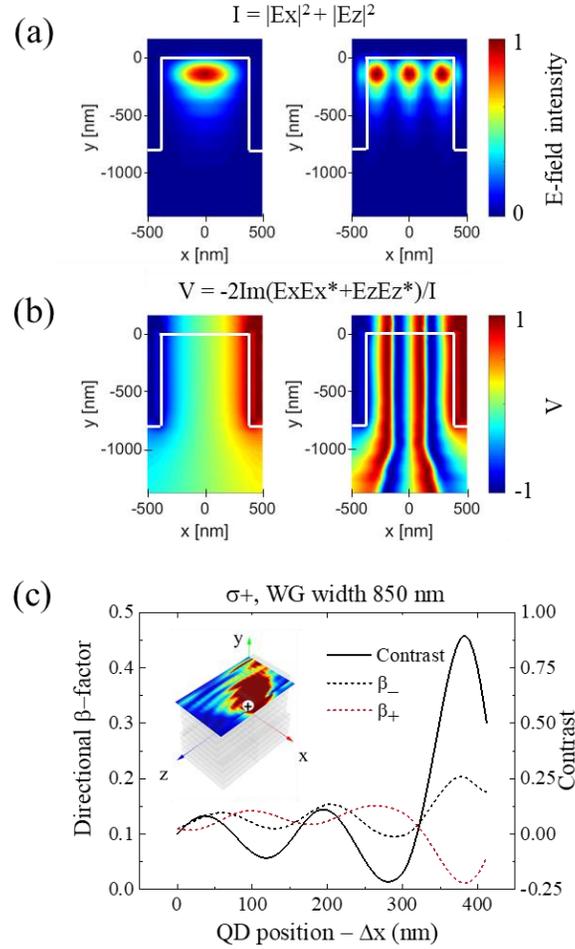

**Figure 4.** Numerical results for a DBR WG structure of width 850 nm and height 800 nm. (a) Calculated intensity distributions for the fundamental and a 3$^{rd}$ order TE-like mode. (b) Calculated distribution of the degree of circular polarization V – see text. (c) Results for a model WG with a dipole emitter in the WG showing directional $\beta$-factors and contrast as a function of dipole displacement $\Delta x$. The inset visualizes a cross-section through the field distribution for the dipole placed at $\Delta x$ = 350 nm.

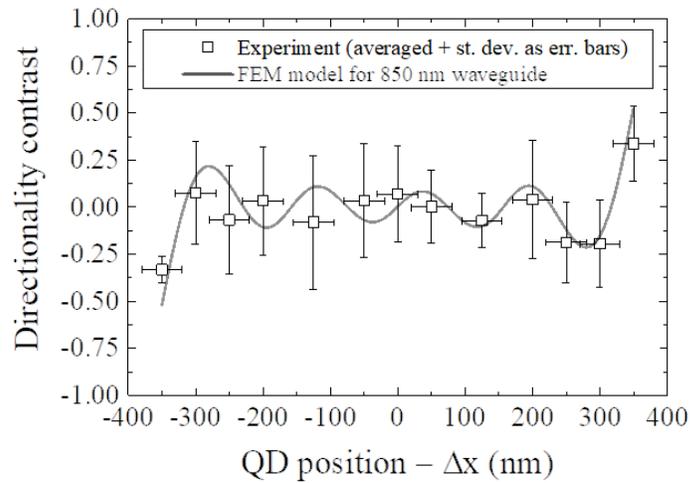

**Figure 5.** Experimental results for the directionality contrast obtained for deterministically fabricated QD-waveguide structures as a function of the QD position Δx together with numerically calculated data for an 850 nm wide waveguide using the FEM method as described in the text.

ASSOCIATED CONTENT

**Supporting Information**.

Finite element method results related to the multiple waveguide propagating modes and directional dipole coupling. (PDF)

AUTHOR INFORMATION

**Corresponding Author**

* email: stephan.reitzenstein@physik.tu-berlin.de

**Author Contributions**

The manuscript was written through contributions of all authors. PM – optical characterization, numerical calculations, manuscript preparation; PS – in-situ EBL, sample processing, manuscript preparation; JS - proximity correction; PG, SB – numerical calculations; AK – sample growth (MOCVD); S. Rodt, S. Reitzenstein – supervising; PM and S. Reitzenstein initiated the research. All authors have checked and given approval to the final version of the manuscript.

**Funding Sources**

Polish Ministry of Science and Higher Education, Mobilność Plus – Vedycja.

The research leading to these results has received funding from the German Research Foundation through CRC 787 'Semiconductor Nanophotonics: Materials, Models, Devices' and from the European Research Council under the European Union's Seventh Framework ERC Grant Agreement No 615613.

ACKNOWLEDGMENT

P. Mrowiński gratefully acknowledge the financial support from the Polish Ministry of Science and Higher Education within "Mobilność Plus" programme.

ABBREVIATIONS

QD quantum dot, WG waveguide, OC outcoupler, DBR distributed Bragg reflector, EBL electron beam lithography, SAM spin-angular momentum, PL photoluminescence, FEM finite element method, SEM scanning electron microscopy, MOCVD metal organic chemical vapor deposition, CL cathodoluminescence

# Supplementary to: Directional emission of a deterministically fabricated quantum dot – Bragg reflection multi-mode waveguide system

Paweł Mrowiński,[1] Peter Schnauber,[1] Philipp Gutsche,[2] Arsenty Kaganskiy,[1] Johannes Schall,[1] Sven Burger,[2] Sven Rodt[1] and Stephan Reitzenstein[1]

[1]Institut für Festkörperphysik, Technische Universität Berlin, Hardenbergstraße 36, 10623 Berlin, Germany

[2]Zuse Institute Berlin, Takustraße 7, 14195 Berlin, Germany

## I. Numerical analysis of directional emission

Calculation of the directional coupling of emission from a circularly polarized transition is performed using the finite-element method (FEM solver JCMsuite). The directional coupling is expressed by the $\beta_{\pm}$-factor which can be evaluated by the overlap integral of the scattered electromagnetic field of the emitter and the field of the guided modes. The eigenmode computation is realized first using a 2D model of the DBR-waveguide system of 850 nm width and 800 nm height, as shown in Fig. 1a), revealing 10 guided modes localized in the top GaAs material (5 TE and 5 TM polarized). Next, in a 3D computational domain we solved time-harmonic Maxwell's equations for a point-like oscillating dipole with circular polarization, simulating a single QD transition from a charged excitonic state. The dipole position is varied from 0 to 400 nm and the field is collected at both waveguide ends at y = +/- 800 nm. In Fig. 1b) a 3D model structure is shown and in Fig. 1c) the calculated emission from the circularly polarized dipole placed at Δx = 350 nm is shown in view perspectives demonstrating highly directional emission along –z direction.

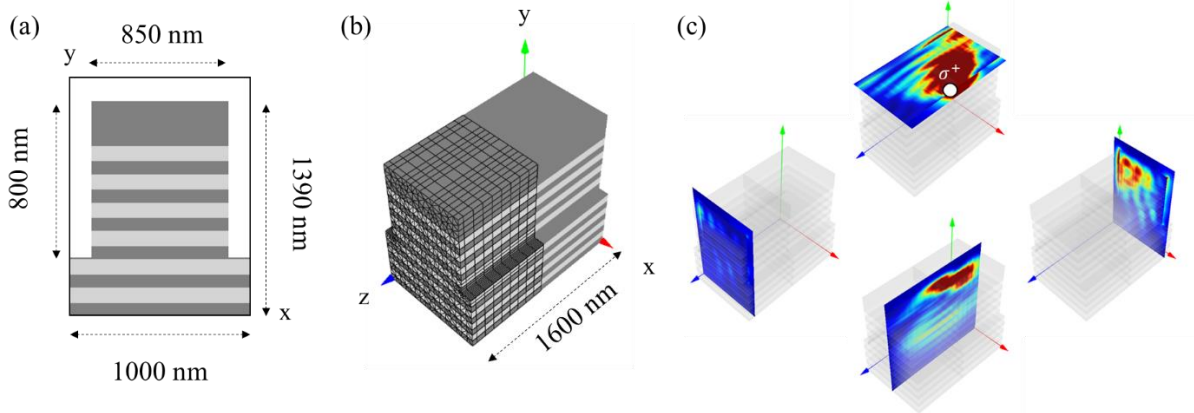

Figure 1 a) 2D model structure of the waveguide consisting of 230 nm of GaAs (dark grey) and GaAs/AlGaAs (dark grey/grey) 8xDBR considered in the FEM calculations of eigenmodes. b) A 3D model structure for calculations of dipole emission and c) four visualization graphs showing the intensity profiles of a circularly polarized dipole located at 350 nm off-center position (white spot).

In Fig. IIa) the calculated TE mode field intensity distributions are shown. The directional β-factor has been evaluated by the overlap of the dipole field at the waveguide ends and the TE mode fields. The dependence on Δx-position of the dipole emitter for the selected TE modes is shown in Fig. 2b) from $\Delta x = 0$ to $\Delta x = 400$ nm. In the following step we evaluated the contrast C, which is presented in Fig. 2c). The contrast presented in the main text in Fig. 5c) is a result of adding all contributions in a way that $C = \frac{\beta'_- - \beta'_+}{\beta'_- + \beta'_+}$, where $\beta'_\pm = \sum_{i=1}^{5} \beta'_{\pm,i}$.

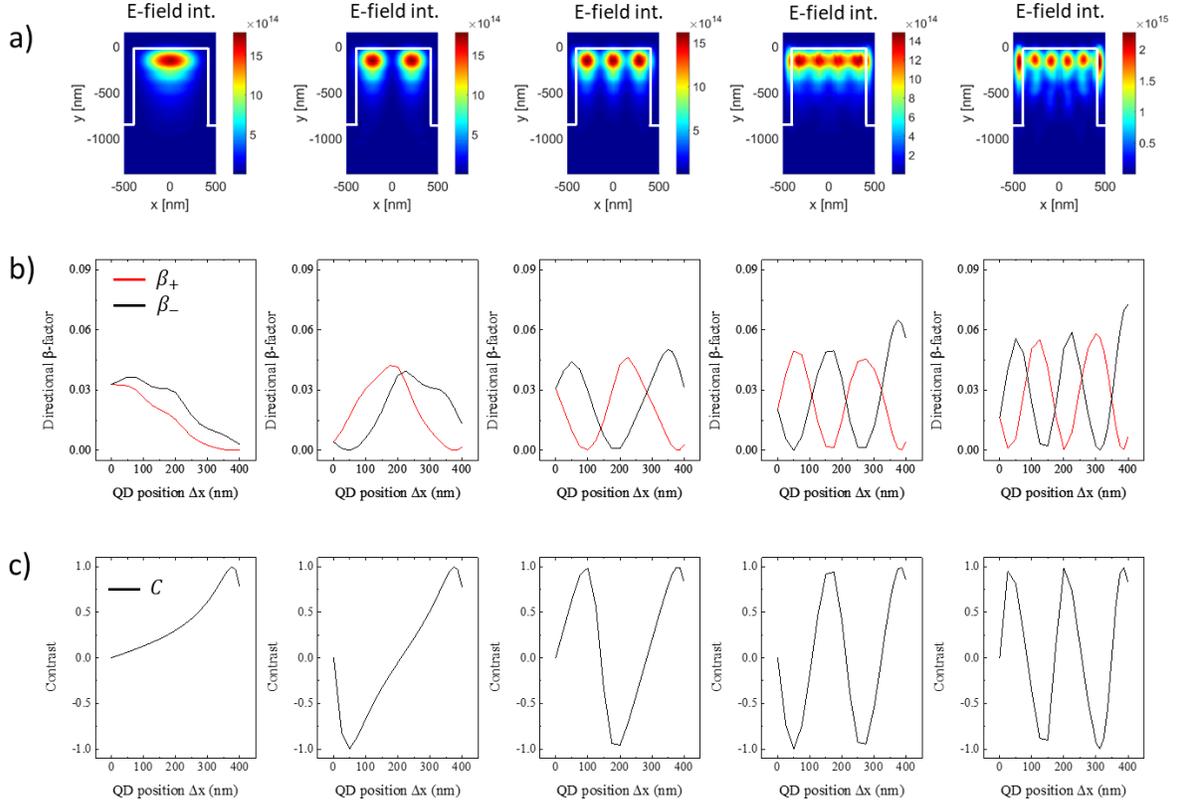

Figure 2 a) TE eigenmodes calculated for the waveguide structures shown in Fig. 1. and directional beta b) and contrast c) in dependence on the QD (dipole) lateral position in the waveguide.

25